\documentclass[showpacs,aps,graphicx,twocolumn]{revtex4}

\usepackage{txfonts}
\usepackage{graphicx}
\usepackage[bookmarks=false]{hyperref}
 \usepackage{ifpdf}

\begin{document}

\title{Efficient single-photon-assisted entanglement concentration for partially entangled photon pairs}

\author{Yu-Bo Sheng,$^{1,2,4}$\footnote{shengyb@njupt.edu.cn} Lan Zhou,$^3$ Sheng-Mei Zhao,$^{1,4}$  and
Bao-Yu Zheng$^{1,4}$}
\address{$^1$ Institute of Signal Processing  Transmission, Nanjing
University of Posts and Telecommunications, Nanjing, 210003,  China\\
$^2$ College of Telecommunications \& Information Engineering,
Nanjing University of Posts and Telecommunications,  Nanjing,
210003, China\\
$^3$ Beijing National Laboratory for Condensed Matter Physics, Institute of Physics,\\
Chinese Academy of Sciences, Beijing 100190, China\\
$^4$ Key Lab of Broadband Wireless Communication and Sensor Network
Technology, Nanjing University of Posts and Telecommunications,
Ministry of Education, Nanjing, 210003, China\\}

\date{\today }

\begin{abstract}
We present two realistic entanglement concentration protocols (ECPs)
for pure partially  entangled photons.  A partially entangled photon
pair can be concentrated to a maximally entangled pair  with only an
ancillary single photon in a certain probability, while the
conventional ones require two copies of partially entangled pairs at
least. Our first protocol is implemented with linear optics and the
second one is implemented with cross-Kerr nonlinearities. Compared
with other ECPs, they do not need to know the accurate coefficients
of the initial state. With linear optics, it is feasible with
current experiment. With cross-Kerr nonlinearities, it does not
require the sophisticated single-photon detectors and can be
repeated to get a higher success probability. Moreover, the second
protocol can get the higher entanglement transformation efficiency
and it maybe the most economical one by far. Meanwhile, both of
protocols are more suitable for multi-photon system concentration,
because they need less operations and classical communications.  All
these advantages make  two protocols be useful in current
long-distance quantum communications.
\end{abstract}
 \maketitle

\section{Introduction}

Entanglement plays an important role in current quantum information
processing, such as quantum computation \cite{computation1}, quantum
key distribution \cite{Ekert91,rmp}, quantum teleportation
\cite{teleportation}, controlled teleportation
\cite{cteleportation}, dense coding \cite{densecoding}, and
quantum-state sharing \cite{QSTS}. In the past ten years, a large
number of experiments have been reported that quantum computation
and quantum communication are more powerful in many aspects than
their classical counterparts. In order to complete such quantum
information processing protocols, the maximally entangled states are
usually required. However, in a practical transmission and storage,
the entanglement inevitably will contact with the environment, and
the noise will make the entanglement degrade. Generally speaking,
the maximally entangled state such as Bell state
$|\phi^{+}\rangle=\frac{1}{\sqrt{2}}
(|H\rangle|H\rangle+|V\rangle|V\rangle)$ may become a mixed state.
That is
\begin{eqnarray}
\rho=F|\phi^{+}\rangle\langle\phi^{+}|+
(1-F)|\psi^{+}\rangle\langle\psi^{+}|.
\end{eqnarray}
Here $|\psi^{+}\rangle=\frac{1}{\sqrt{2}}
(|H\rangle|V\rangle+|V\rangle|H\rangle)$.
 It also may become a partially entangled
state
$|\varphi\rangle=\alpha|H\rangle|H\rangle+\beta|V\rangle|V\rangle$,
with $|\alpha|^{2}+|\beta|^{2}=1$. Here $|H\rangle$ ($|V\rangle$)
represents the horizontal  (vertical) photon polarization.

Entanglement purification can distill a set of mixed entangled
states into a subset of highly entangled states with local operation
and classical communication
\cite{Bennett1,Deutsch,Pan1,Pan2,Murao,shengpra,Horodecki,Yong}.
However, it can only  improve the quality of the mixed state and can
not get the maximally entangled state. On the other hand,
entanglement concentration, which will be detailed can be used to
convert the partially entangled pairs to the maximally entangled
ones
\cite{Bennett2,swapping1,swapping2,Yamamoto1,Yamamoto2,zhao1,zhao2,wangxb,kim,shengpra2,shengqic,shengpla}.
In 1996, Bennett \emph{et al.} proposed the first entanglement
concentration protocol (ECP), named Schmidt decomposition protocol.
In their protocol, they use the collective measurements which are
difficult to manipulate in experiment \cite{Bennett2}. Bose \emph{et
al.} also proposed an ECP based on entanglement swapping
\cite{swapping1}. But their protocol requires the collective
Bell-state measurement. Moreover, they need to know the coefficients
in order to reconstruct the same entangled states. In 2001, Zhao
\emph{et al.} and Yamamoto \emph{et al.} proposed two similar ECPs
independently with polarization beam splitters (PBSs)
\cite{zhao1,Yamamoto1,zhao2,Yamamoto2}. They have simplified the
schmidt projection method and adopted the parity check to substitute
the original collective measurement.  Here we call it PBS1 protocol.
This idea has  been developed to reconstruct the ECP with cross-Kerr
nonlinearity, which can be used to construct the quantum
nondemolition detectors (QND) \cite{shengpra2}. Here we call it QND1
protocol.

The current ECPs have in common that, for instance, in Refs.
\cite{zhao1,wangxb,kim,shengpra2,Yamamoto1,shengqic,shengpla} they
need at least two copies of less-entangled  (or  say partially
entangled) pairs initially. But after performing the protocols, at
most  one pair of maximally entangled state can be obtained, or both
of each should be discarded, according to the measurement results by
classical communication.  Local operation and classical
communication can not be used to increase entanglement. Therefor,
these ECPs are essentially  the transformation of entanglement and
the previous works of entanglement concentration are not optimal.
Actually, two copies  of less-entangled pairs are not necessary.
Using one copy of less-entangled pair to distillate high quality
entanglement has been proposed in continuous variables system. In
Ref. \cite{singlecopy1}, Opatrn\'{y} \emph{et al.} showed the
improvement on teleportation of continuous variables by photon
subtraction via conditional measurement. In 2008, He and Bergou
proposed a general probabilistic approach for transforming a single
copy of discrete entanglement state without classical communication
\cite{singlecopy2}.

In this paper, we describe two single-photon-assisted ECPs  in which
only one pair of less-entangled state and one single photon are
required. The two ECPs are focused on the practical discrete
less-entangled photon pairs and they are implemented with  linear
optics and cross-Kerr nonlinearity, respectively. Comparing with
current ECPs, the single-photon-assisted ECPs are more economical.
In the first ECP, with linear optics, it can reach the same success
probability as the protocol of Ref. \cite{zhao1}, but requires only
one pair of less-entangled state.  In the second ECP, with the help
of cross-Kerr nonlinearity, it can be repeated to get a higher
success probability. These advantages make the two protocols more
feasible in practical applications.

This paper is organized as follows: in Sec.II, we first explain the
basic principle of  single-photon-assisted ECP with linear optics.
We call it PBS2 protocol. In Sec.III, we  extend this protocol to
the system of cross-Kerr nonlinearity. We call it QND2 protocol. We
show that sophisticated single-photon detectors are not required and
the discarded items in PBS2 protocol can also be reused to perform
concentration.  The higher success probability can be obtained than
other protocols. In Sec. IV, we first calculate the entanglement
transformation efficiency, and  then
 make a discussion and summary.

\section{single-photon-assisted entanglement concentration with linear optics}

The basic principle of our PBS2 protocol is shown in Fig.1.  The
less-entangled pair of photons emitted from $S_{1}$ are sent to
Alice and Bob. The photon $a$ belongs to Alice and $b$ belongs to
Bob. The initial photon pair is in the following unknown state:
\begin{eqnarray}
|\Phi\rangle_{a1b1}=\alpha|H\rangle_{a1}|H\rangle_{b1}+\beta|V\rangle_{a1}|V\rangle_{b1}.\label{lessentangled}
\end{eqnarray}
The another source $S_{2}$ emits a single photon with the form of
\begin{eqnarray}
|\Phi\rangle_{a2}=\alpha|H\rangle_{a2}+\beta|V\rangle_{a2}.\label{singlephoton}
\end{eqnarray}
Here $|\alpha|^{2}+|\beta|^{2}=1$. $a1$, $b1$, and $a2$ are
different spatial modes.

\begin{figure}
\begin{center}
\includegraphics[width=8cm,angle=0]{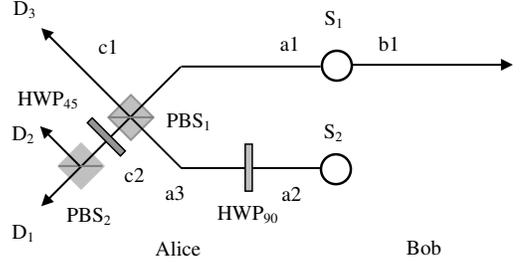}
\caption{A schematic drawing of the single-photon-assisted ECPs with
linear optics. $S_{1}$ is the partial entanglement source and
$S_{2}$ is the single photon source. PBSs transmit the horizontal
polarization component and reflect the vertical component.
HWP$_{90}$ and HWP$_{45}$ can rotate the polarization of the state
by $90^{\circ}$ and $45^{\circ}$, respectively.}
\end{center}
\end{figure}

The initial state of the three photons can be written as:
\begin{eqnarray}
|\Psi\rangle&=&|\Phi\rangle_{a1b1}\otimes|\Phi\rangle_{a2}\nonumber\\
&=&\alpha^{2}|H\rangle_{a1}|H\rangle_{b1}|H\rangle_{a2}+\alpha\beta|H\rangle_{a1}|H\rangle_{b1}|V\rangle_{a2}\nonumber\\
&+&\alpha\beta|V\rangle_{a1}|V\rangle_{b1}|H\rangle_{a2}+\beta^{2}|V\rangle_{a1}|V\rangle_{b1}|V\rangle_{a2}.
\end{eqnarray}
 Alice first rotates the polarization state of  the single photon
$|\Phi\rangle_{a2}$ by $90^{\circ}$ by half-wave plate (HWP$_{90}$
in Fig.1). Then the state can be rewritten as:
\begin{eqnarray}
|\Psi\rangle'&=&|\Phi\rangle_{a1b1}\otimes|\Phi'\rangle_{a3}\nonumber\\
&=&\alpha^{2}|H\rangle_{a1}|H\rangle_{b1}|V\rangle_{a3}+\alpha\beta|H\rangle_{a1}|H\rangle_{b1}|H\rangle_{a3}\nonumber\\
&+&\alpha\beta|V\rangle_{a1}|V\rangle_{b1}|V\rangle_{a3}+\beta^{2}|V\rangle_{a1}|V\rangle_{b1}|H\rangle_{a3}\nonumber\\
&=&\alpha^{2}|H\rangle_{a1}|V\rangle_{a3}|H\rangle_{b1}+\beta^{2}|V\rangle_{a1}|H\rangle_{a3}|V\rangle_{b1}\nonumber\\
&+&\alpha\beta
(|H\rangle_{a1}|H\rangle_{a3}|H\rangle_{b1}+|V\rangle_{a1}|V\rangle_{a3}|V\rangle_{b1}).\label{system}
\end{eqnarray}
From above equation, it is evident that the items
$|H\rangle_{a1}|H\rangle_{a3}|H\rangle_{b1}$ and
$|V\rangle_{a1}|V\rangle_{a3}|V\rangle_{b1}$ will lead the two
output modes $c1$ and $c2$ both exactly contain only one  photon.
However, item $|H\rangle_{a1}|V\rangle_{a3}|H\rangle_{b1}$ will lead
two photons both in $c2$ mode, and item
$|V\rangle_{a1}|H\rangle_{a3}|V\rangle_{b1}$ will lead both photons
in $c1$ mode. Therefor, by choosing the three-mode cases, i.e. each
modes of $c1$, $c2$ and $b1$  exactly contain and only contain one
photon, the initial state is projected into a maximally three-photon
entangled state:
\begin{eqnarray}
|\Psi\rangle''=\frac{1}{\sqrt{2}}
(|H\rangle_{c1}|H\rangle_{c2}|H\rangle_{b1}+|V\rangle_{c1}|V\rangle_{c2}|V\rangle_{b1}),\label{threephoton}
\end{eqnarray}
with a probability of $2|\alpha\beta|^{2}$.

In order to generate a maximally entangled Bell-state between Alice
and Bob, they could perform a $45^{\circ}$ polarization measurement
onto the photon $c2$. In Fig.1, with quarter-wave plate (HWP$_{45}$
in Fig.1), it can make:
\begin{eqnarray}
|H\rangle_{c2}\rightarrow\frac{1}{\sqrt{2}} (|H\rangle_{c2}+|V\rangle_{c2}),\nonumber\\
|V\rangle_{c2}\rightarrow\frac{1}{\sqrt{2}}
(|H\rangle_{c2}-|V\rangle_{c2}).
\end{eqnarray}

After the rotation, Eq. (\ref{threephoton}) will evolve to
\begin{eqnarray}
|\Psi\rangle'''=\frac{1}{2} (|H\rangle_{c1}|H\rangle_{b1}+|V\rangle_{c1}|V\rangle_{b1})|H\rangle_{c2}\nonumber\\
+
(|H\rangle_{c1}|H\rangle_{b1}-|V\rangle_{c1}|V\rangle_{b1})|V\rangle_{c2}.\label{threephoton2}
\end{eqnarray}
Now Alice lets the photon $c2$ pass through the PBS$_{2}$. Clearly,
if the detector $D_{1}$ fires, the photon pair is left in the state
as
\begin{eqnarray}
|\phi^{+}\rangle_{a1b1}=\frac{1}{\sqrt{2}}
(|H\rangle_{c1}|H\rangle_{b1}+|V\rangle_{c1}|V\rangle_{b1}).\label{max1}
\end{eqnarray}

If the detector $D_{2}$ fires,  the photon pair is left in the state
as
\begin{eqnarray}
|\phi^{-}\rangle_{a1b1}=\frac{1}{\sqrt{2}}
(|H\rangle_{c1}|H\rangle_{b1}-|V\rangle_{c1}|V\rangle_{b1}).\label{max2}
\end{eqnarray}
Both of Eqs. (\ref{max1}) and  (\ref{max2}) are the maximally
entangled states. One of them say Alice or Bob only needs to perform
a phase flip to convert Eq. (\ref{max2}) to  (\ref{max1}), and the
whole concentration process is finished. It has the  success
probability of $2|\alpha\beta|^{2}$, which is the same as Ref.
\cite{zhao1}.  During the whole process, they do not require two
copies of less-entangled pair, and only one pair and a single photon
are required. Meanwhile, they  do not need to know the exact
coefficients of the initial states $|\Phi\rangle_{a1b1}$ and
$|\Phi\rangle_{a2}$, but only require them to equal.

From Fig.1, it is straightforward to extend this protocol to
reconstruct maximally entangled multi-partite Greenberg-
Horne-Zeilinger  (GHZ) states from partially entangled GHZ states.

The partially multi-partite entangled GHZ state with $N$ photons can
be written as:
\begin{eqnarray}
|\Phi\rangle_{N}=\alpha|H\rangle_{A}H\rangle_{B}\cdots
H\rangle+\beta|V\rangle_{A}V\rangle_{B}\cdots V\rangle.
\end{eqnarray}
Here the subscriptions $A$, $B\cdots$ denote the parties Alice, Bob,
Charlie etc.. Each one owns one photon. With the same principle
described above, Alice needs another single photon with the same
form of Eq. (\ref{singlephoton}).  After passing through the PBS, if
the output modes of Alice's PBS both contain exactly one photon,
then the whole system collapses into a $N+1$ maximally entangled
state. Then with the same principle, they can obtain the $N$-photon
maximally entangled state by measuring one photon after rotating
$45^{\circ}$.

Interestingly, this protocol seems more feasibly for the
concentration of multi-photon GHZ system. First, in this protocol,
only Alice needs to perform this protocol, while in the conventional
protocols \cite{zhao1,shengpra2,Yamamoto1}, all  the parties should
perform the same operations with Alice. Second, in this protocol,
only Alice asks other parities to retain or discard their photons,
and they do not need to check their measurement results. It greatly
simplifies the complication of classical communication. Third,
multi-photon GHZ states are more difficult to generate in current
condition. We can have the same success probability but only  need
one pair of less-entangled GHZ state.

\section{single-photon-assisted entanglement concentration with cross-Kerr nonlinearity}

So far, we have briefly  described the single-photon-assisted ECP
 based on linear optics. In  above description, Alice
exploits the PBS and sophisticated single-photon detectors to
distinguish $|HH\rangle$ and $|VV\rangle$ from  $|HV\rangle$ and
$|VH\rangle$. It is essentially the parity-check measurement of
polarization photons. Unfortunately, with current technology,
sophisticated single-photon detectors are not likely to be
available, that makes this protocol cannot be achieved simply with
only linear optics. In this section, we will introduce the
cross-Kerr nonlinearity to construct QND, which can also be used to
implement this protocol. Cross-kerr nonlinearity has been widely
studied in the construction of CNOT gate \cite{QND1}, performance of
complete Bell-state analysis \cite{QND2}, entanglement purification
\cite{shengpra} and so on
\cite{lin1,lin2,lin3,he1,he2,he3,he4,he5,shengbellstateanalysis}.
The Hamiltonian of a cross-Kerr nonlinear medium can be written as
$H=\hbar\chi \hat{n_{a}}\hat{n_{b}}$. Here the $\hbar\chi$ is the
coupling strength of the nonlinearity, which is decided by the
material of cross-Kerr. The $ \hat{n_{a}} (\hat{n_{b}})$ is the
number operator for mode $a (b)$.

\begin{figure}
\begin{center}
\includegraphics[width=6cm,angle=0]{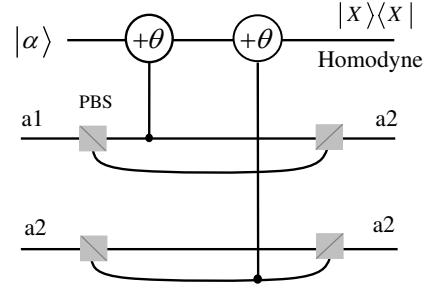}
\caption{The schematic drawing of the principle of the quantum
nondemolition detector (QND) constructed by the cross-Kerr
nonlinearity. It can also be shown in Ref. \cite{shengpra2}. One can
distinguish the state $|HH\rangle$ and $|VV\rangle$ from
$|HV\rangle$ and $|VH\rangle$ by different phase shift of the
coherent state.}
\end{center}
\end{figure}

If we consider a coherent state $|\alpha\rangle$ and a single photon
$|\psi\rangle=\gamma|0\rangle+\delta|1\rangle$ interact with the
cross-Kerr nonlinearity, the whole system can be described as
\cite{QND1,QND2}:
\begin{eqnarray}
U_{ck}|\psi|\alpha\rangle&=& (\gamma|0\rangle+\delta|1\rangle)|\alpha\rangle\nonumber\\
&=&\gamma|0\rangle|\alpha\rangle+\delta|1\rangle|\alpha
e^{i\theta}\rangle.
\end{eqnarray}
Here $|0\rangle$ and $|1\rangle$ are the Fock states, which mean no
photon and one photon respectively. $\theta=\chi t$ and $t$ is the
interaction time. It is obvious that the phase shift of coherent
state is directly proportional to the number of photons.

In Fig.3, we adopt the QND to substitute the PBS. Then the whole
system $|\Phi\rangle_{a1b1}\otimes|\Phi'\rangle_{a3}$ with the
coherent state $|\alpha\rangle$ can be written as:
\begin{eqnarray}
&&|\Psi\rangle'|\alpha\rangle=|\Phi\rangle_{a1b1}\otimes|\Phi'\rangle_{a3}|\alpha\rangle\nonumber\\
&=& (\alpha^{2}|H\rangle_{a1}|H\rangle_{b1}|V\rangle_{a3}+\alpha\beta|H\rangle_{a1}|H\rangle_{b1}|H\rangle_{a3}\nonumber\\
&+&\alpha\beta|V\rangle_{a1}|V\rangle_{b1}|V\rangle_{a3}+\beta^{2}|V\rangle_{a1}|V\rangle_{b1}|H\rangle_{a3})|\alpha\rangle\nonumber\\
&\rightarrow&\alpha^{2}|H\rangle_{a1}|V\rangle_{a3}|H\rangle_{b1}|\alpha e^{i2\theta}\rangle+\beta^{2}|V\rangle_{a1}|H\rangle_{a3}|V\rangle_{b1}|\alpha\rangle\nonumber\\
&+&\alpha\beta
(|H\rangle_{a1}|H\rangle_{a3}|H\rangle_{b1}+|V\rangle_{a1}|V\rangle_{a3}|V\rangle_{b1})|\alpha
e^{i\theta}\rangle.\nonumber\\\label{system2}
\end{eqnarray}
In above evolution, the items $|H\rangle_{a1}|V\rangle_{a3}$ and
$|V\rangle_{a1}|H\rangle_{a3}$ pick up the phase shift with
$2\theta$ and no phase shift, respectively. But the items
$|H\rangle_{a1}|H\rangle_{a3}$ and $|V\rangle_{a1}|V\rangle_{a3}$
both pick up the phase shift with $\theta$. Therefor, if the phase
shift of homodyne measurement is $\theta$, Alice asks Bob to keep
the whole state. Otherwise they discard the state. The remained
state essentially is the state described in Eq. (\ref{threephoton}).
Therefor, following the same step described above, one can
ultimately obtain the maximally entangled state
$|\phi^{+}\rangle_{c1b1}$ if $D_{1}$ fires, and get
$|\phi^{-}\rangle_{c1b1}$ if $D_{2}$ fires.

\begin{figure}
\begin{center}
\includegraphics[width=8cm,angle=0]{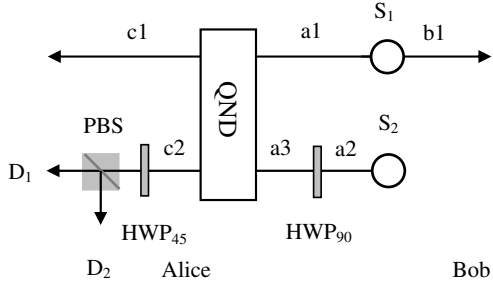}
\caption{The schematic drawing of the ECP with QND. Here we
substitute the PBS shown in Fig.1 with QND. This ECP can obtain a
higher success probability.}
\end{center}
\end{figure}

In  above description, Alice only picks up the instance in which the
phase shift $\theta$ on his coherent state, and discards the other
instances.  If a suitable cross-Kerr medium is available and Alice
can control the interaction time $t$ exactly, which makes the phase
shift $\theta=\pi$, in this way, one can not distinguish the phase
shift $0$ and $2\pi$. The discarded items in the above equation are
\begin{eqnarray}
|\Phi'\rangle=\alpha^{2}|H\rangle_{c1}|V\rangle_{a3}|H\rangle_{b1}+\beta^{2}|V\rangle_{c1}|H\rangle_{a3}|V\rangle_{b1},\label{moreless}
\end{eqnarray}
with the probability of $|\alpha|^{4}+|\beta|^{4}$.  Alice uses the
HWP$_{45}$ to rotate the photon in $c2$ and finally it is detected
by $D_{1}$ or $D_{2}$. Eq. (\ref{moreless}) becomes
\begin{eqnarray}
|\Phi''\rangle=\alpha^{2}|H\rangle_{c1}|H\rangle_{b1}+\beta^{2}|V\rangle_{c1}|V\rangle_{b1},\label{moreless1}
\end{eqnarray}
 if $D_{1}$ fires, and becomes
\begin{eqnarray}
|\Phi'''\rangle=\alpha^{2}|H\rangle_{c1}|H\rangle_{b1}-\beta^{2}|V\rangle_{c1}|V\rangle_{b1},\label{moreless2}
\end{eqnarray}
if $D_{2}$ fires.

$|\Phi''\rangle$ and $|\Phi'''\rangle$ are both the partially
entangled states which can also be used to reconcentrate to a
maximally entangled states. For instance, if they get
$|\Phi''\rangle$, Alice only needs to choose another single photon
with the form of $\alpha^{2}|H\rangle_{a2}+\beta^{2}|V\rangle_{a2}$,
and follows the same method described above. That is:
\begin{eqnarray}
&& (\alpha^{2}|H\rangle_{c1}|H\rangle_{b1}+\beta^{2}|V\rangle_{c1}|V\rangle_{b1})\nonumber\\
&\otimes& (\alpha^{2}|H\rangle_{a2}+\beta^{2}|V\rangle_{a2})\otimes|\alpha\rangle\nonumber\\
&\rightarrow& (\alpha^{2}|H\rangle_{c1}|H\rangle_{b1}+\beta^{2}|V\rangle_{c1}|V\rangle_{b1})\nonumber\\
&\otimes& (\alpha^{2}|V\rangle_{a2}+\beta^{2}|H\rangle_{a2})\otimes|\alpha\rangle\nonumber\\
&\rightarrow&\alpha^{4}|H\rangle_{a1}|V\rangle_{a3}|H\rangle_{b1}|\alpha e^{i2\theta}\rangle+\beta^{4}|V\rangle_{a1}|H\rangle_{a3}|V\rangle_{b1}|\alpha\rangle\nonumber\\
&+& (\alpha\beta)^{2}
(|H\rangle_{a1}|H\rangle_{a3}|H\rangle_{b1}+|V\rangle_{a1}|V\rangle_{a3}|V\rangle_{b1})|\alpha
e^{i\theta}\rangle.\nonumber\\\label{system3}
\end{eqnarray}
Alice can also pick up the phase shift $\theta$ and get the same
state as Eq. (\ref{threephoton}). The success probability of
obtaining Eq. (\ref{threephoton}) is
\begin{eqnarray}
P_{2}=\frac{2|\alpha\beta|^{4}}{|\alpha|^{4}+|\beta|^{4}}.
\end{eqnarray}

We can also get
\begin{eqnarray}
P_{3}&=&\frac{2|\alpha\beta|^{8}}{|\alpha|^{8}+|\beta|^{8}},\nonumber\\
 &\cdots&\nonumber\\
P_{N}&=&\frac{2|\alpha\beta|^{2^{N}}}{|\alpha|^{2^{N}}+|\beta|^{2^{N}}},
\end{eqnarray}
where $N$ is the iteration number of our concentration processes.
\begin{figure}
\begin{center}
\includegraphics[width=7cm,angle=0]{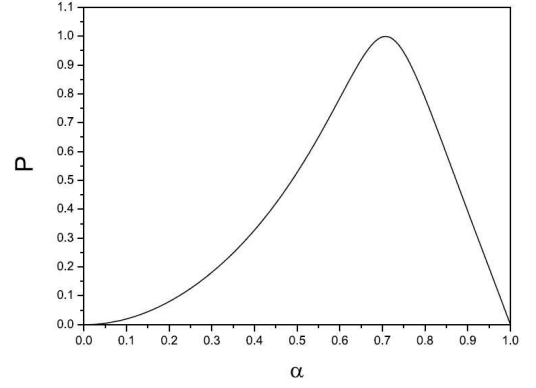}
\caption{The success probability P of getting a maximally entangled
state after performing the QND2 protocol $N$ times
 ($N\rightarrow\infty$) is altered with the entanglement of the
initial partially entangled state, i.e. $\alpha$. For numerical
simulation, we let $N=10$ as a good approximation.}
\end{center}
\end{figure}

The total success probability to get a maximally entangled state
from the initial  partially entangled state  is
\begin{eqnarray}
P=P_{1}+P_{2}+\cdots+P_{N}=\sum_{N=1}^{\infty}P_{N}.
\end{eqnarray}
Interestingly, if $\alpha=\beta=\frac{1}{\sqrt{2}}$,
$P=\frac{1}{2}+\frac{1}{4}+\cdots+\frac{1}{2^{N}}=1$. But if
$\alpha\neq\beta$, $P<1$.
 Fig. 4 shows that the relationship
between the coefficient of initial partially entangled state
$\alpha$ and total success probability $P$. From Fig. 4, it is shown
that the success probability is not a fixed value. It is related
with the  entanglement of the initial state, and it increases with
the entanglement of the initial partially entangled state.

We should point out that during a practical operation,
  the longer interaction time will induce decoherence from
losses. It will make the output state become a mixed one. Therefor,
by controlling the longer interaction time to make the phase shift
$\theta=\pi$ may not seem like an efficient way. Fortunately,
 another better alternative way is to rotate the coherent state in
Eq. (\ref{system2}) by $\theta$. After rotation,  Eq.
(\ref{system2}) becomes
\begin{eqnarray}
&\rightarrow&\alpha^{2}|H\rangle_{a1}|V\rangle_{a3}|H\rangle_{b1}|\alpha e^{i\theta}\rangle\nonumber\\
&+&\beta^{2}|V\rangle_{a1}|H\rangle_{a3}|V\rangle_{b1}|\alpha e^{-i\theta}\rangle\nonumber\\
&+&\alpha\beta
(|H\rangle_{a1}|H\rangle_{a3}|H\rangle_{b1}+|V\rangle_{a1}|V\rangle_{a3}|V\rangle_{b1})|\alpha\rangle.\label{system3}
\end{eqnarray}
From above description, if the coherent state picks up no phase
shift, the remained state is also the same as Eq.
(\ref{threephoton}). Otherwise, one can use the $|X\rangle\langle
X|$ homodyne detection \cite{QND1}, which makes the $|\alpha e^{\pm
i\theta}\rangle$ can not be distinguished. In this way, the
discarded state is also the same as it is described in Eq.
(\ref{moreless}). Moreover, with the help of QND, this protocol can
also be extended to multi-photon system, and be used to reconstruct
maximally entangled multi-photon GHZ state. It has the same success
probability as shown in Fig.4.

\section{discussion and summary}

By far, we have fully described our protocols with both PBS and QND.
In each protocol, we only require one pair of less-entangled photons
and a single photon. It is known that local operation and classical
communication can not increase the entanglement. Therefor,
entanglement concentration is essentially the transformation of
entanglement. We  define the entanglement transformation efficiency
$\eta$ as:
\begin{eqnarray}
\eta=\frac{E_{c}}{E_{0}}.
\end{eqnarray}
Here $E_{0}$ is the entanglement of  initial partially entangled
state, and $E_{c}$ is the entanglement of the state after performing
concentration one time. $E_{c}$ can be described as:
\begin{eqnarray}
E_{c}=P_{s}\times1+ (1-P_{s})\times E'.\label{Ec}
\end{eqnarray}

The first item of Eq. (\ref{Ec}) means that after concentration, we
 get the maximally entangled state with the success probability
$P_{s}$. The second item means that the concentration is failure,
and we get a more less-entangled pair. Obviously, if we use the PBS
to perform the concentration, the second item is 0 for it collapses
to a separated  state $|HV\rangle$ or $|VH\rangle$ in different
spatial modes \cite{zhao1,Yamamoto1}. For two-body pure entangled
state, Von Neumann entropy is suitable to describe the entanglement.
Therefor, the entanglement of the initial state in Eq.
(\ref{lessentangled}) can be described as:
\begin{eqnarray}
 E=-|\alpha|^{2}\log_{2}|\alpha|^{2}-|\beta|^{2}\log_{2}|\beta|^{2}.
\end{eqnarray}

\begin{figure}
\begin{center}
\includegraphics[width=7cm,angle=0]{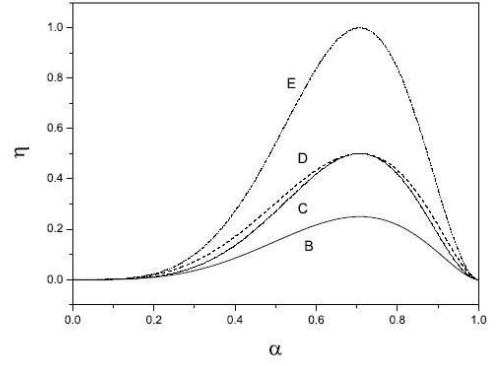}
\caption{The entanglement transformation efficiency $\eta$  is
altered with  the coefficient $\alpha$  after performing each
protocols one time. Curves B, C, D and E correspond to the protocols
of PBS1 \cite{zhao1}, QND1 \cite{shengpra2}, PBS2 and QND2
respectively.  All the curves show that $\eta$ increases with the
entanglement of the initial entangled state. The QND2 protocol has
the highest transformation efficiency. It can reach maximum value 1
when $\alpha=\frac{1}{\sqrt{2}}$.}
\end{center}
\end{figure}

We calculate the $\eta$ of PBS1 protocol as \cite{zhao1}:
\begin{eqnarray}
\eta_{PBS1}=\frac{2|\alpha\beta|^{2}\times1}{2E}=\frac{|\alpha\beta|}{E}.
\end{eqnarray}
The '2' in the denominator means that initially we need two copies
of less-entangled states with entanglement $E$.

For QND1 protocol \cite{shengpra2},
\begin{eqnarray}
\eta_{QND1}=\frac{E'_{QND1}}{2E},
\end{eqnarray}
with
\begin{eqnarray}
E'_{QND1}&=&2|\alpha\beta|^{2}\nonumber\\
&+& (|\alpha|^{4}+|\beta|^{4})[-\frac{|\alpha|^{4}}{|\alpha|^{4}
+|\beta|^{4}}\log_{2}\frac{|\alpha|^{4}}{|\alpha|^{4}+|\beta|^{4}}\nonumber\\
&-&\frac{|\beta|^{4}}{|\alpha|^{4}
+|\beta|^{4}}\log_{2}\frac{|\beta|^{4}}{|\alpha|^{4}+|\beta|^{4}}].
\end{eqnarray}
\begin{figure}
\begin{center}
\includegraphics[width=7cm,angle=0]{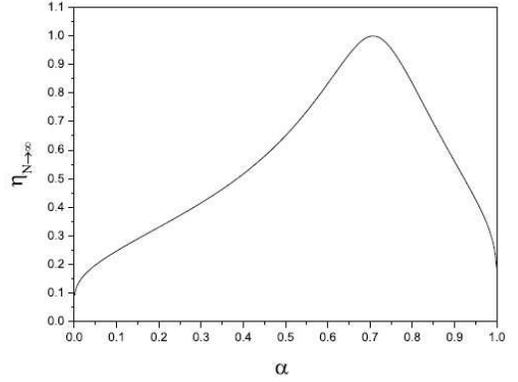}
\caption{The entanglement transformation efficiency $\eta$ plotted
against $\alpha$ after performing each protocols $N$ times
($N\rightarrow\infty$) in QND2 protocol. For numerical simulation,
we let $N=10$ as a good approximation.}
\end{center}
\end{figure}

In our protocol, we only need one pair of less-entangled state to
perform the protocol. Therefor, in PBS2 protocol,
\begin{eqnarray}
\eta_{PBS2}=\frac{2|\alpha\beta|^{2}}{E}=2\eta_{PBS1},
\end{eqnarray}
and in QND2 protocol,
\begin{eqnarray}
\eta_{QND2}=\frac{E'_{QND1}}{E}=2\eta_{QND1}.
\end{eqnarray}
The relationship between the coefficient $\alpha$ and entanglement
transformation efficiency  is shown in Fig.5. It is shown that
$\eta$ is also not a fixed value, but increases with the initial
entanglement. In QND2 protocol, $\eta$ can reach the max value 1
with $\alpha=\frac{1}{\sqrt{2}}$, which means that the initial one
is the maximally entangled state.  But in traditional protocols
\cite{zhao1,shengpra2}, $\eta\leq0.5$.

 We also calculate
the limit of   entanglement transformation  efficiency of QND2
protocol, by iterating the protocol $N$ ($N\rightarrow\infty$)
times.
\begin{eqnarray}
\eta^{N\rightarrow\infty}_{QND2}=\frac{\sum_{N=1}^{\infty}E_{N}P_{N}}{E_{0}}=\frac{P}{E_{0}}.
\end{eqnarray}
The $E_{N}$ means the entanglement of remained state after
performing successful concentration in Nth iteration. It is a
maximally entangled state with $E_{N}=1$. Fig.6 shows the
relationship between $\alpha$ and transformation probability $\eta$.
Obviously, $\eta$ is monotone increasing with the entanglement of
the initial state, and can get the max value 1 when the initial
state is maximally entangled one, as $\alpha=\frac{1}{\sqrt{2}}$.

In this paper, the basic elements for us to complete the task are
the PBS and QND. In fact, both of them act as the same role, that is
parity check. In Refs \cite{zhao1} and  \cite{shengpra2}, they also
resort to PBS and QND to perform the concentration.  But in each
step, they require two pairs of less-entangled states. Our protocol
shows that with only one pair of less-entangled state and a single
photon, we can also achieve this task. This good feature makes these
protocols have a higher entanglement transformation efficiency than
others. In the processing of describing our concentration protocol,
we exploit entanglement source and single photon source. An idea
single photon source should exactly emit one and only one photon
when the device is triggered.  However, no single photon source and
entanglement source will be ideal.  In current technology, practical
pulse generated by a source may contain no photons or multiple
photons, with different probability. We denote $P_{m}$ as the
probability of emitting $m$ photons. Interestingly, $P_{0}$ means no
photon and will not give rise to errors. It only decreases the
success probability of the protocol. Because  only two photons can
not satisfy the three-mode cases. However, $m\geq2$ will give rise
to errors. For example, $m=2$ will lead both of the modes $c1$ and
$c2$ contain one photon in Fig.1 which is a success event in our
protocol. Fortunately, it is possible to make the probability of
such events rather small. In Ref. \cite{singlephoton}, it was
reported that a single photon source whose $P_{0}$  and $P_{2}$ are
14\% and 0.08\% respectively. Current spontaneous parametric
down-conversion entanglement source is analogy with the single
photon source. It generates entangled pair with the form of
\cite{Yamamoto1}
\begin{eqnarray}
|\Upsilon\rangle=\sqrt{g}
(|vac\rangle+\gamma|\phi^{+}\rangle+\gamma^{2}|\phi^{+}\rangle^{2}+\cdots).
\end{eqnarray}

The multi-photon items $|\phi^{+}\rangle^{2}$ can also cause errors.
In practical teleportation experiments, the $\gamma^{2}\sim10^{-4}$
\cite{teleportationexperiment1,teleportationexperiment2}, and the
errors of multi-photon items is negligible.

In Sec.III, we exploit cross-Kerr nonlinearity to implement our
protocol. Although a lot of works have been studied in the area of
cross-Kerr nonlinearity
\cite{QND1,QND2,lin1,lin2,lin3,he1,he2,he3,he4,he5,shengbellstateanalysis,shengpra2,shengqic},
we should acknowledge that it is still a quite  controversial
assumption for a clean cross-Kerr nonlinearity in the optical
single-photon regime. In 2002, Kok \emph{et al.} pointed out that
the Kerr phase shift is only $\tau\approx10^{-18}$ in the optical
single-photon regime \cite{kok1,kok2}. In 2003, Hofmann showed that
with a single two-level atom in a one-sided cavity, a large
phase-shift of $\pi$ can be achieved \cite{hofmann}. Gea-Banacloche
 showed that large shifts via the giant Kerr effect with
single-photon wave packet is impossible in current technology
\cite{Gea}. The results of previous works of Shapiro and Razavi are
consistent with Gea-Banacloche \cite{Shapiro1,Shapiro2}. Recently,
He \emph{et al.} discussed the feasibility of QND relies on the
compatibility of small phase shift with large coherent state
amplitude. They developed a general theory of the iteration between
continuous-mode photonic pulses  and applied it to the case of
single photon interacting with a coherent state. They showed that if
the pulses can fully pass through each other and the unwanted
transverse-mode effects can be suppressed, the high fidelities,
nonzero conditional phases, and high photon numbers are compatible
\cite{he3}. Recent research also showed that using weak measurement,
it is possible to amplify a cross-Kerr phase shift to an observable
value, which is much larger than the intrinsic magnitude of the
single-photon-level nonlinearity \cite{meaurement}.

In summary, we present two different protocols for nonlocal
entanglement concentration of partially entangled states. We exploit
both PBS and cross-Kerr nonlinearity to achieve the task. Our
protocols have several advantages: first, they do not need to know
the exactly efficiency $\alpha$ and $\beta$ of the less-entangled
pairs. Second, they also do not resort to the collective
measurement. Third, with QND, it does not require the parties to
adopt the sophisticated single-photon detectors, and it can be
iterated to get a higher success probability. Fourth, compared with
the previous works, the most significant advantage
 is that in each step we only need one pair of
less-entangled state.  It provides our protocols to obtain more
higher entanglement transformation efficiency  than  others. Fifth,
these protocols are more feasible for multi-photon GHZ state
concentration, because they greatly reduce the practical operations
and simplify the complication of classical communication for each
parties.  All these advantages may make our protocols be more useful
in practical applications.

\section*{ACKNOWLEDGEMENTS}

This work is supported by the National Natural Science Foundation of
China under Grant No. 11104159,  Scientific Research Foundation of
Nanjing University of Posts and Telecommunications under Grant No.
NY211008,  University Natural Science Research Foundation of JiangSu
Province under Grant No. 11KJA510002,  the open research fund of Key
Lab of Broadband Wireless Communication and Sensor Network
Technology  (Nanjing University of Posts and Telecommunications),
Ministry of Education, China, and a Project Funded by the Priority
Academic Program Development of Jiangsu Higher Education
Institutions.

\end{document}